\documentclass[aps,preprint,showpacs,showkeys]{revtex4}%
\usepackage{amsfonts}
\usepackage{amsmath}
\usepackage{amssymb}
\usepackage{graphicx}%
\providecommand{\U}[1]{\protect\rule{.1in}{.1in}}

\begin{document}
\preprint{ }
\title{A genuine four-partite entangled state}
\author{Ming-Yong Ye}
\email{myye@mail.ustc.edu.cn}
\affiliation{School of Physics and Optoelectronics Technology, Fujian Normal University,
Fuzhou 350007, People's Republic of China}
\author{Xiu-Min Lin}
\affiliation{School of Physics and Optoelectronics Technology, Fujian Normal University,
Fuzhou 350007, People's Republic of China}

\begin{abstract}
In a recent paper, a genuine four-partite entangled state is proposed [Y. Yeo
and W. K. Chua, Phys. Rev. Lett. \textbf{96}, 060502 (2006)], which has been
found to have many interesting entanglement properties. We show this state is
locally equivalent to some graph states.

\end{abstract}
\maketitle

Quantum entanglement has many applications such as teleportation \cite{tele}
and dense coding \cite{dense}. Recently Yeo and Chua found a genuine
four-qubit entangled state $\left\vert \chi^{00}\right\rangle $ that can be
used to teleport an arbitrary two-qubit state \cite{four}. The state
$\left\vert \chi^{00}\right\rangle $ proposed in Ref. \cite{four} has the form%
\begin{equation}
\left\vert \chi^{00}\right\rangle =\frac{1}{\sqrt{2}}\left(  \left\vert
\zeta^{0}\right\rangle +\left\vert \zeta^{1}\right\rangle \right)
_{A_{3}A_{4}B_{1}B_{2}},
\end{equation}
with%
\begin{align}
\left\vert \zeta^{0}\right\rangle  &  =\frac{1}{2}\left(  \left\vert
0000\right\rangle -\left\vert 0011\right\rangle -\left\vert 0101\right\rangle
+\left\vert 0110\right\rangle \right)  ,\\
\left\vert \zeta^{1}\right\rangle  &  =\frac{1}{2}\left(  \left\vert
1001\right\rangle +\left\vert 1010\right\rangle +\left\vert 1100\right\rangle
+\left\vert 1111\right\rangle \right)  .
\end{align}
It has been found that $\left\vert \chi^{00}\right\rangle $ has many
interesting entanglement properties \cite{four}. It has maximal entanglement
between $A_{3}A_{4}$ and $B_{1}B_{2}$ or between $A_{3}B_{1}$ and $A_{4}B_{2}%
$, while it is not reducible to a pair of Bell states. The amount of
entanglement between $A_{3}B_{2}$ and $A_{4}B_{1}$ has the value $1$ when we
use von Neumann entropy as an entanglement measure. It has properties that
differ from those of four-party GHZ and W states \cite{four}. Due to its
interesting entanglement properties, the state $\left\vert \chi^{00}%
\right\rangle $ attracts much attention \cite{in,chen,man}.

It is not clear why the state $\left\vert \chi^{00}\right\rangle $ has so many
interesting entanglement properties. We will show $\left\vert \chi
^{00}\right\rangle $ is locally equivalent to some graph states. First let us
give a simple introduction to graph states \cite{graph}. Let $G=\left(
V,E\right)  $ be a graph, in which $V$ is the set of vertices and $E$ is the
set of edges. The graph state $\left\vert G\right\rangle $ corresponding to
the graph $G$ is the pure state
\begin{equation}
\left\vert G\right\rangle =%
{\displaystyle\prod\limits_{\left\{  a,b\right\}  \in E}}
U_{ab}\left\vert +\right\rangle ^{V},
\end{equation}
where $\left\vert +\right\rangle =\left(  \left\vert 0\right\rangle
+\left\vert 1\right\rangle \right)  /\sqrt{2}$ and $U_{ab}$ is the
controlled-phase gate%
\begin{equation}
U_{ab}=\left\vert 0\right\rangle _{a}\left\langle 0\right\vert _{a}\otimes
I^{b}+\left\vert 1\right\rangle _{a}\left\langle 1\right\vert _{a}%
\otimes\sigma_{z}^{b}.
\end{equation}
For an instance, the graph $G_{a}$ in FIG. 1(a) corresponds to the graph
state
\begin{equation}
\left\vert G_{a}\right\rangle =U_{A_{3}A_{4}}U_{B_{1}B_{2}}U_{A_{3}B_{1}%
}U_{A_{4}B_{2}}\left\vert +\right\rangle _{A_{3}}\left\vert +\right\rangle
_{A_{4}}\left\vert +\right\rangle _{B_{1}}\left\vert +\right\rangle _{B_{2}}.
\end{equation}
Two graphs $G=\left(  V,E\right)  $ and $G^{^{\prime}}=\left(  V,E^{^{\prime}%
}\right)  $ are locally equivalent, if there is a local unitary $U\in U\left(
2\right)  ^{V}$ such that their corresponding graph states%
\begin{equation}
\left\vert G^{^{\prime}}\right\rangle =U\left\vert G\right\rangle .
\end{equation}
Local equivalence of graphs is connected to a graph transformation called
local complementation: suppose $G=\left(  V,E\right)  $ is a graph and $a\in
V$, the local complement of $G$ at $a$, denoted by $\tau_{a}\left(  G\right)
$, is a new graph obtained by complementing the neighborhood of $a$ and
leaving the rest of the graph unchanged. It has been proved that two graphs
$G=\left(  V,E\right)  $ and $G^{^{\prime}}=\left(  V,E^{^{\prime}}\right)  $
will be locally equivalent if they are related by a sequence of local
complementations, i.e., $G^{^{\prime}}=\tau_{a_{1}}\circ\ldots\circ\tau
_{a_{n}}\left(  G\right)  $ for some $a_{1},\ldots,a_{n}\in V$
\cite{graph,add}. It can be easily found that the graph $G_{a}$ in FIG. 1(a)
and the graph $G_{b}$ in FIG. 1(b) are locally equivalent because they are
connected by a local complementation:%
\begin{equation}
G_{b}=\tau_{A_{4}}\left(  G_{a}\right)  .
\end{equation}
The local equivalence between $G_{a}$ and $G_{b}$ can be confirmed by the
relation between their corresponding graph states:%
\begin{equation}
\left\vert G_{b}\right\rangle =e^{i\frac{\pi}{4}}e^{-i\frac{\pi}{4}\sigma
_{z}^{A_{3}}}e^{i\frac{\pi}{4}\sigma_{x}^{A_{4}}}e^{-i\frac{\pi}{4}\sigma
_{z}^{B_{2}}}\left\vert G_{a}\right\rangle . \label{x1}%
\end{equation}
Actually all graphs in FIG. 1 are locally equivalent since they are connected
by local complementations. For more about graph states we refer to Ref.
\cite{graph} and the references therein.

Our goal is to show $\left\vert \chi^{00}\right\rangle $ is locally equivalent
to graph states depicted by graphs in FIG. 1. It can be easily checked that
$\left\vert \chi^{00}\right\rangle $ is locally equivalent to $\left\vert
G_{b}\right\rangle $:%
\begin{equation}
\left\vert \chi^{00}\right\rangle =\sigma_{z}^{A_{3}}\sigma_{z}^{B_{2}%
}H^{B_{2}}\left\vert G_{b}\right\rangle , \label{x2}%
\end{equation}
where $H^{B_{2}}$ is the Hadamard gate acting on qubit $B_{2}$. Because graph
states depicted in FIG. 1 are locally equivalent, the state $\left\vert
\chi^{00}\right\rangle $ is locally equivalent to all these graph states.
Since locally equivalent states have the same entanglement properties, the
entanglement properties of $\left\vert \chi^{00}\right\rangle $ can be
obtained by studying the graph state $\left\vert G_{a}\right\rangle $ or other
locally equivalent graph states depicted in FIG. 1. It is not hard to find
that $\left\vert G_{a}\right\rangle $ has all the above mentioned properties
of $\left\vert \chi^{00}\right\rangle $.

Now we give two more properties about $\left\vert \chi^{00}\right\rangle $. It
is known that graph states are also stabilizer states \cite{graph}, and the
graph state $\left\vert G_{b}\right\rangle $ is stabilized by the following
four independent commuting observables:
\begin{align}
K_{1}  &  =\sigma_{x}^{A_{3}}\sigma_{z}^{A_{4}}\sigma_{z}^{B_{1}}\sigma
_{z}^{B_{2}},\\
K_{2}  &  =\sigma_{z}^{A_{3}}\sigma_{x}^{A_{4}}\sigma_{z}^{B_{2}},\\
K_{3}  &  =\sigma_{z}^{A_{3}}\sigma_{x}^{B_{1}}\sigma_{z}^{B_{2}},\\
K_{4}  &  =\sigma_{z}^{A_{3}}\sigma_{z}^{A_{4}}\sigma_{z}^{B_{1}}\sigma
_{x}^{B_{2}}.
\end{align}
From Eq. (\ref{x2}) we know $\left\vert \chi^{00}\right\rangle $ is also a
stabilizer state and it is stabilized by
\begin{align}
\bar{K}_{1}  &  =\sigma_{z}^{A_{3}}\sigma_{z}^{B_{2}}H^{B_{2}}K_{1}\sigma
_{z}^{A_{3}}H^{B_{2}}\sigma_{z}^{B_{2}}=\sigma_{x}^{A_{3}}\sigma_{z}^{A_{4}%
}\sigma_{z}^{B_{1}}\sigma_{x}^{B_{2}},\\
\bar{K}_{2}  &  =\sigma_{z}^{A_{3}}\sigma_{z}^{B_{2}}H^{B_{2}}K_{2}\sigma
_{z}^{A_{3}}H^{B_{2}}\sigma_{z}^{B_{2}}=-\sigma_{z}^{A_{3}}\sigma_{x}^{A_{4}%
}\sigma_{x}^{B_{2}},\\
\bar{K}_{3}  &  =\sigma_{z}^{A_{3}}\sigma_{z}^{B_{2}}H^{B_{2}}K_{3}\sigma
_{z}^{A_{3}}H^{B_{2}}\sigma_{z}^{B_{2}}=-\sigma_{z}^{A_{3}}\sigma_{x}^{B_{1}%
}\sigma_{x}^{B_{2}},\\
\bar{K}_{4}  &  =\sigma_{z}^{A_{3}}\sigma_{z}^{B_{2}}H^{B_{2}}K_{4}\sigma
_{z}^{A_{3}}H^{B_{2}}\sigma_{z}^{B_{2}}=\sigma_{z}^{A_{3}}\sigma_{z}^{A_{4}%
}\sigma_{z}^{B_{1}}\sigma_{z}^{B_{2}},
\end{align}
that is to say%
\begin{equation}
\bar{K}_{i}\left\vert \chi^{00}\right\rangle =\left\vert \chi^{00}%
\right\rangle ,i=1,\ldots,4. \label{xa}%
\end{equation}
Eq. (\ref{xa}) can be directly checked from the expressions of $\left\vert
\chi^{00}\right\rangle $ and $\bar{K}_{i}$.

Another property about $\left\vert \chi^{00}\right\rangle $ is that it also
admits a GHZ-argument \cite{ghz}. According to observables $\bar{K}_{1}$,
$\bar{K}_{2}$, $\bar{K}_{4}$ and
\begin{equation}
\bar{K}_{1}\bar{K}_{2}\bar{K}_{4}=\sigma_{x}^{A_{3}}\sigma_{x}^{A_{4}}%
\sigma_{z}^{B_{2}},
\end{equation}
we can design four different measurement settings. In each measurement
setting, the measurement results on different qubits will have\ a correlation.
The observables $\bar{K}_{1}$, $\bar{K}_{2}$, $\bar{K}_{4}$ and $\bar{K}%
_{1}\bar{K}_{2}\bar{K}_{4}$ lead to the following correlations respectively
\begin{align}
m_{x}^{A_{3}}m_{z}^{A_{4}}m_{z}^{B_{1}}m_{x}^{B_{2}}  &  =1,\label{q1}\\
-m_{z}^{A_{3}}m_{x}^{A_{4}}m_{x}^{B_{2}}  &  =1,\label{q2}\\
m_{z}^{A_{3}}m_{z}^{A_{4}}m_{z}^{B_{1}}m_{z}^{B_{2}}  &  =1,\label{q3}\\
m_{x}^{A_{3}}m_{x}^{A_{4}}m_{z}^{B_{2}}  &  =1, \label{q4}%
\end{align}
where $m_{x}^{A_{3}}=\pm1$ and $m_{z}^{A_{3}}=\pm1$ denote measurement
outcomes if the qubit $A_{3}$ is measured in spin-$x$ or spin-$z$ direction.
In local hidden variable theories all these four Eqs. (\ref{q1}-\ref{q4})
should be satisfied at the same time. However the product of Eq. (\ref{q1}),
(\ref{q2}) and (\ref{q3}) gives%
\begin{equation}
m_{x}^{A_{3}}m_{x}^{A_{4}}m_{z}^{B_{2}}=-1,
\end{equation}
which contradict with Eq. (\ref{q4}), so local hidden variable theories cannot
be used to explain the measurement results on state $\left\vert \chi
^{00}\right\rangle $.

In conclusion we have shown that the interesting four-partite genuine
entangled state $\left\vert \chi^{00}\right\rangle $ proposed in \cite{four}
is locally equivalent to graph states depicted by graphs in FIG. 1. This
finding uncovers the mysteries of the state $\left\vert \chi^{00}\right\rangle
$ and gives an explanation why it has so many interesting entanglement
properties. We have also shown $\left\vert \chi^{00}\right\rangle $ is a
stabilizer state and there is a GHZ-argument for $\left\vert \chi
^{00}\right\rangle $.

This work was funded by National Natural Science Foundation of China (Grant
No. 10574022), the Fujian Provincial Natural Science Foundation (Grant No.
A0410016 and No. 2006J0230), and the Foundation for Universities in Fujian
Province (Grant No. 2007F5041).

\end{document}